\documentclass[journal=jctcce,manuscript=article]{achemso}
\setkeys{acs}{maxauthors=0,articletitle=true}
\usepackage{achemso}
\usepackage{placeins}
\usepackage{graphics}
\usepackage{amssymb,amsfonts}
\usepackage{graphicx}
\usepackage[table,dvipsnames]{xcolor}
\usepackage{multirow}
\usepackage{caption}
\usepackage{subcaption}
\usepackage{booktabs}
\usepackage{colortbl}
\usepackage{amsmath}
\usepackage{amsopn}
\usepackage{bm}
\usepackage{siunitx}
\usepackage{bm}
\usepackage{color}
\usepackage{array}
\usepackage{lscape}
\usepackage{mciteplus}
\usepackage[version=3]{mhchem}
\usepackage{ulem}
\usepackage{listings}
\usepackage{enumerate}
\usepackage{lmodern}

\author{Frederik {\O}. Kjeldal}
\affiliation{DTU Chemistry, Technical University of Denmark\\Kemitorvet Bldg. 206, 2800 Kgs. Lyngby, Denmark}
\author{Janus J. Eriksen}
\email{janus@dtu.dk}
\affiliation{DTU Chemistry, Technical University of Denmark\\Kemitorvet Bldg. 206, 2800 Kgs. Lyngby, Denmark}

\title[TITLE]{Properties of Local Electronic Structures}

\begin{document}

\begin{abstract}

The simulation of intrinsic contributions to molecular properties holds the potential to allow for chemistry to be directly inferred from changes to electronic structures at the atomic level. In the present study, we demonstrate how such local properties can be readily derived from suitable molecular orbitals to yield effective fingerprints of various types of atoms in organic molecules. In contrast, corresponding inferences from schemes that instead make use of individual atomic orbitals for this purpose are generally found to fail in expressing much uniqueness in atomic environments. By further studying the extent to which entire chemical reactions may be decomposed into meaningful and continuously evolving atomic contributions, schemes based on molecular rather than atomic orbitals are once again found to be the more consistent, even allowing for intricate differences between seemingly uniform nucleophilic substitutions to be probed.

\end{abstract}

\newpage

\section{Introduction}\label{intro_sect}

Among trained synthetic chemists, meticulous labours over the past centuries have helped shape what may be termed as chemical {\textit{Fingerspitzengefühl}}, that is, the educated ability of experienced professionals to gauge property contributions from individual entities within molecules as well as what rules the signs and magnitudes of these~\cite{blanksby_ellison_bde_acr_2003}. Be these comprised of individual atoms, bonds, or even entire functional moieties, if one desires to devise practical and functional mappings between theory and experiments it is instructive to study how successfully such compositional features may be numerically inferred from standard computational simulations. However, the act of discerning what precisely underpins a reaction or a molecular transformation solely on the basis of a simulation of its electronic structure at large potentially risks conflating any changes to this at a local level~\cite{ruedenberg_chem_bond_rmp_1962,shaik_chem_bond_ccr_2017}. Without ways of associating reactivities, solubilities, or redox potentials with intrinsic molecular attributes, it is hard to imagine how to advance on today's rational engineering of such chemical features~\cite{aspuru_guzik_lindh_reiher_acs_cent_sci_2018}.\\

To that end, a myriad of computational schemes have been proposed over the years for identifying local components of globally defined electronic structures, e.g., as predicted by Kohn-Sham density functional theory~\cite{hohenberg1964inhomogeneous,kohn1965self,parr_yang_dft_book} (KS-DFT). In the course of the present work, we wish to compare two classes of schemes~\cite{nakai_eda_partitioning_cpl_2002,nakai_eda_partitioning_ijqc_2009,eriksen_decodense_jcp_2020,eriksen_local_condensed_phase_jpcl_2021,eriksen_elec_ex_decomp_jcp_2022}, which---on the basis of either a set of atomic or molecular orbitals (AOs or MOs)---decompose standard mean-field simulations of electronic structure into individual atomic contributions without recourse to further manipulations or additional calculations~\cite{kitaura_morokuma_eda_ijqc_1976,szalewicz_sapt_chem_rev_1994,hohenstein_sherrill_sapt_wires_2012,head_gordon_eda_pnas_2017,head_gordon_eda_avpc_2021,frenking_eda_wires_2012,frenking_eda_wires_2018}. Importantly, we will not be explicitly concerned with bonds, the many types of these, nor their assumed properties~\cite{grimme_chem_bond_jacs_1996,schleyer_chem_bond_jpca_2001,weinhold_nbo_book_2005,weinhold_nbo_review_wires_2012,weinhold_nbo_review_jcc_2012,rahm_hoffmann_chem_bond_jacs_2015,rahm_hoffmann_chem_bond_jacs_2016}, but rather with the characteristic manners in which an atomic environment and its local electronic structure can be thought of as implicitly defined by the network of bonds that surround said atom, substituent groups and their inductive or mesomeric effects, different structural motifs, and so on.\\

By studying decompositions of molecules in their equilibrium geometries, but also how the atomic properties derived from these vary along well-defined reaction coordinates, we will explore to what extent AO- and MO-based partitioning schemes are capable of yielding unique fingerprints of distinct local environments around atoms embedded within molecules. In addition, we will also be concerned with the possible ways such derived properties align with chemical intuition. However, just as ambiguities necessarily surround the infinitely many ways by which to partition a quantum-mechanical observable into the components of a many-body system, so will a universal consensus on the definition of some of the key concepts in chemistry also not prevail among all chemists, e.g., on electronegativity~\cite{pauling_electronegativity_jacs_1932,mulliken_electronegativity_jcp_1934,mulliken_electronegativity_jcp_1935}, possible exceptions to the octet rule~\cite{lewis_valence_book_1923,musher_hypervalency_angew_chem_1969}, or even the nature of bonding itself~\cite{frenking_bonding_chem_rev_2019}. By definition, the formation of a thermodynamically stable molecule from its atomic constituents is associated with a lowering of the molecular energy over the sum of the corresponding free-atom energies. When viewed through a decomposition scheme, this atomization energy may be interpreted as a sum of atomic contributions, each of which reflect the response of said atom to molecular embedding. Herein, focus is on whether the contribution of an atom to a molecular energy decreases or increases upon embedding within a given molecule (indicating a net stabilization or destabilization, respectively), how great this response is, and how to rationalize relative changes when modifying, for instance, hybridization and oxidation state.\\

In recent work of ours, we applied a number of AO- and MO-based partitioning schemes across a small model of chemical compound space, consisting of less than 10k constitutional and structural molecular isomers each limited to a mere 7 heavy atoms (C, N, O, and S)~\cite{eriksen_qm7_ml_jctc_2023}. Its small extent notwithstanding, this model space still comprised a number of fundamentally distinct local atomic environments and our work was concerned, in part, with the study of how atomic contributions to cohesive energies were distributed across the space, but also how the aggregate and diversity in the representation of each element from these decompositions influenced the rate by which atomic contributions could be learned using various kinds of machine learning architectures based on neural networks. In the course of the present work, we will instead seek to quantify what factors fundamentally determine the magnitude and sign of atomic contributions to molecular atomization energies. Given how the electronic structure in the vicinity of a nucleus is necessarily perturbed upon shifting frames from a vacuum to a molecular environment, we will inspect apparent (de)stabilizations to C, N, O, and S atoms in molecules on the basis of nuclear-electronic interactions shared between different partitioning schemes. Results for a broad selection of typical organic compounds are presented. Next, we will build on these trends to study how local energies and dipole moments change along chemical reaction coordinates, with an initial application to simple concerted S$_\mathrm{N}$2 substitution reactions involving a halomethane and a halide of the same kind. 

\section{Methods}\label{methods_sect}

In the present work, the total KS-DFT energy of any given molecule consisting of $\mathcal{M}$ atoms will be expressed on the following, somewhat unconventional form:
\begin{align}
E = \sum^{\mathcal{M}}_{K}E^{\text{nuc}}_{K} + \bar{E}^{\text{el}}_{K} = \sum^{\mathcal{M}}_{K}E^{\text{nuc}}_{K} + E^{\text{el}}_{K} \ . \label{atom_decomp_eq}
\end{align}

The individual contributions to Eq. \ref{atom_decomp_eq} are associated with either the nuclear (nuc) or electronic (el) interactions between an atom $K$ and both its own local environment and its surroundings. Among these, the nuclear terms ($E^{\text{nuc}}$) are shared between both decomposition schemes to be studied here, while the electronic terms are instead inferred on the basis of either the AOs ($\bar{E}^{\text{el}}$) or MOs ($E^{\text{el}}$) of a system at hand. Differences in the electronic contributions associated with unique atoms embedded within a molecule thus give rise to contrasts between general AO- and MO-based schemes~\cite{nakai_eda_partitioning_cpl_2002,nakai_eda_partitioning_ijqc_2009,eriksen_decodense_jcp_2020,eriksen_local_condensed_phase_jpcl_2021,eriksen_elec_ex_decomp_jcp_2022}. Detailed expressions for the components of Eq. \ref{atom_decomp_eq} are presented in Eqs. S1--S4 of the supporting information (SI), as are corresponding expressions for analogous decompositions of molecular dipole moments (Eqs. S5--S7).\\

The two chosen decomposition schemes thus depend either solely on the spatial locality of a set of underlying AOs ($\bar{E}^{\text{el}}$)---denoted as the energy density analysis (EDA) scheme by Nakai~\cite{nakai_eda_partitioning_cpl_2002,nakai_eda_partitioning_ijqc_2009}---or the spatial arrangement of a suitable set of localized linear combinations of AOs ($E^{\text{el}}$), as proposed by Eriksen in Ref. \citenum{eriksen_decodense_jcp_2020}. In the course of the present study, we will directly compare these schemes on the basis of how they yield fingerprints of distinctive atomic environments insofar as they both share $E^{\text{nuc}}$ in common. Any differences between AO- and MO-based partitioning schemes, and how they decompose cohesive energies into atomic contributions, will then be caused by the manners in which they account for the responses of local electronic structures around individual atoms. Important in the present context is the fact that, however potentially informative the proposed sub-decompositions in Eqs. S1--S4 of the SI may be, for instance, in cases where the two schemes yield either similar or vastly incomparable results, these are still most optimally compared on the basis of total atomic contributions as these remain the only unambiguously derived fingerprints.

\section{Computational Details}

In the MO-based decomposition scheme of Ref. \citenum{eriksen_decodense_jcp_2020}, a spatially localized MO basis is used throughout---spanned by intrinsic bonds orbitals (IBOs)---and Mulliken populations derived in a minimal, free-atom intrinsic AO (IAO) basis are used as the weights in Eq. S3 of the SI~\cite{knizia_iao_ibo_jctc_2013}. In the calculations of atomization energies to follow (Figs. \ref{fig_1}--\ref{fig_4}), the B3LYP functional~\cite{becke_b3lyp_functional_jcp_1993,frisch_b3lyp_functional_jpc_1994} is used in combination with the def2svp basis set~\cite{weigend_ahlrichs_def2svp_basis_pccp_2005}. In the subsequent simulations of S$_\mathrm{N}$2 substitutions (Figs. \ref{fig_5}--\ref{fig_7}), however, the range-separated $\omega$B97M-V functional~\cite{mardirossian_head_gordon_wb97m_v_functional_jcp_2016} with VV10 nonlocal correlation~\cite{vydrov_voorhis_vv10_functional_jcp_2010} is used alongside the augmented aug-pc-2 basis set~\cite{jensen_pc_basis_sets_jcp_2001} to warrant a proper treatment of ion-dipole interactions in these reactions. Corresponding results computed without diffuse functions (pc-2) are presented in the SI as Figs. S1--S3.

\section{Results and Discussion}

\begin{figure}[ht!]
    \centering
    \includegraphics[width=\textwidth]{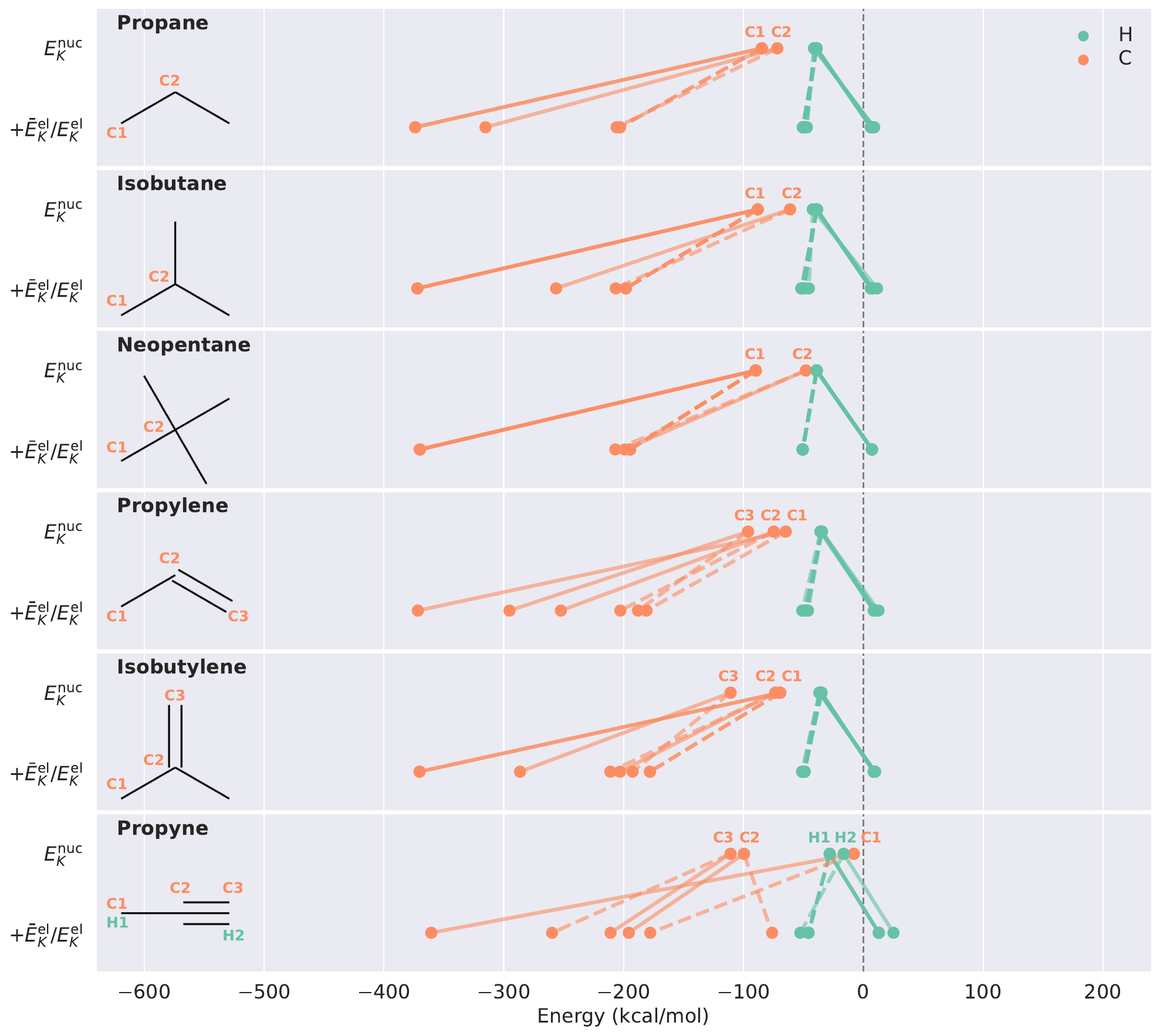}
    \caption{Accumulated nuclear and electronic contributions to atomization energies (cf. Eq. \ref{atom_decomp_eq}) for a selection of alkanes, alkenes, and alkynes. Individual results obtained using the IBO/IAO and EDA schemes are represented by solid and dashed lines, respectively.}
    \label{fig_1}
\end{figure}
\begin{figure}[ht!]
    \centering
    \includegraphics[width=\textwidth]{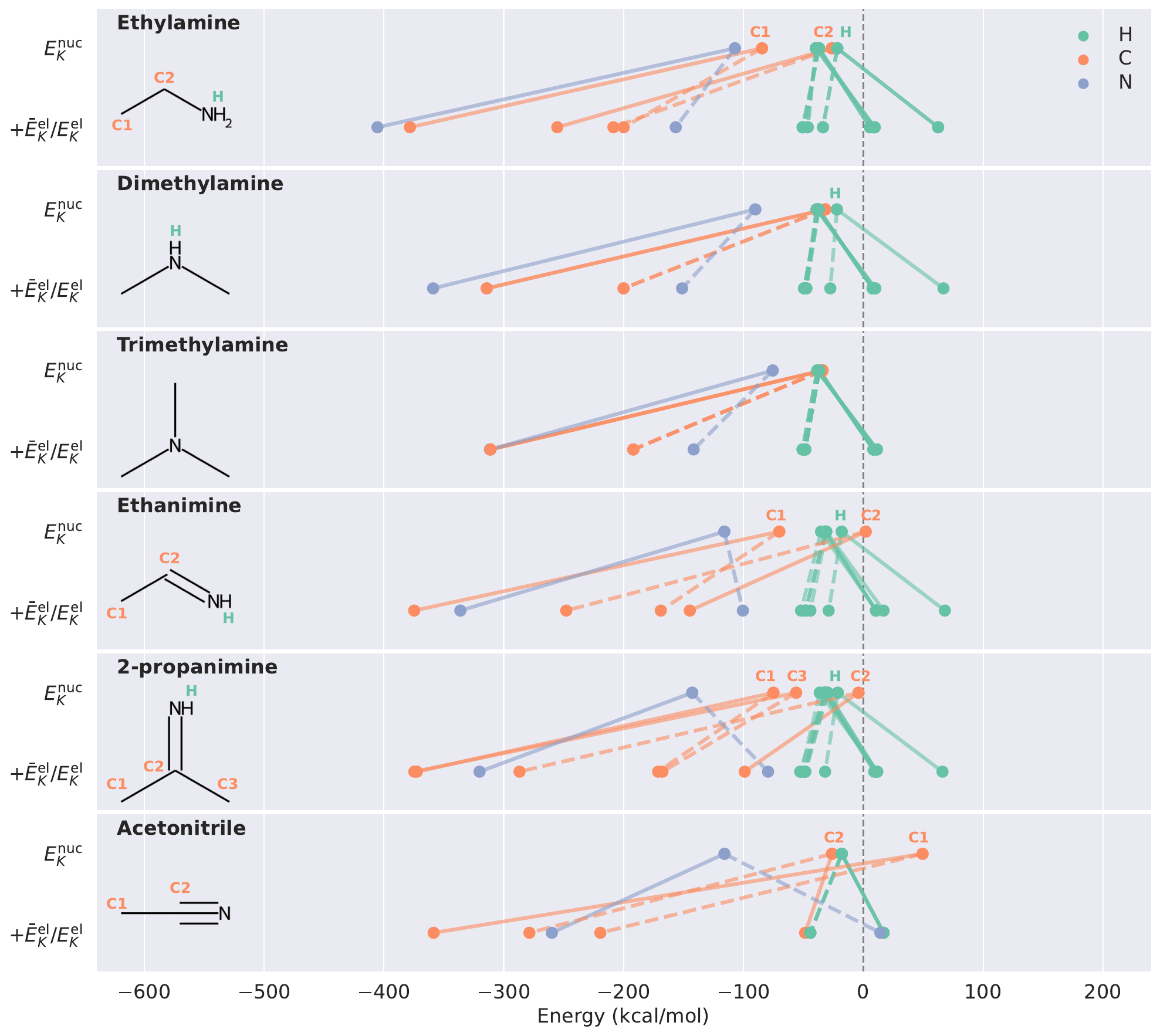}
    \caption{IBO/IAO and EDA results for selected amines, imines, and nitriles. For details, please see the caption to Fig. \ref{fig_1}.}
    \label{fig_2}
\end{figure}
\begin{figure}[ht!]
    \centering
    \includegraphics[width=\textwidth]{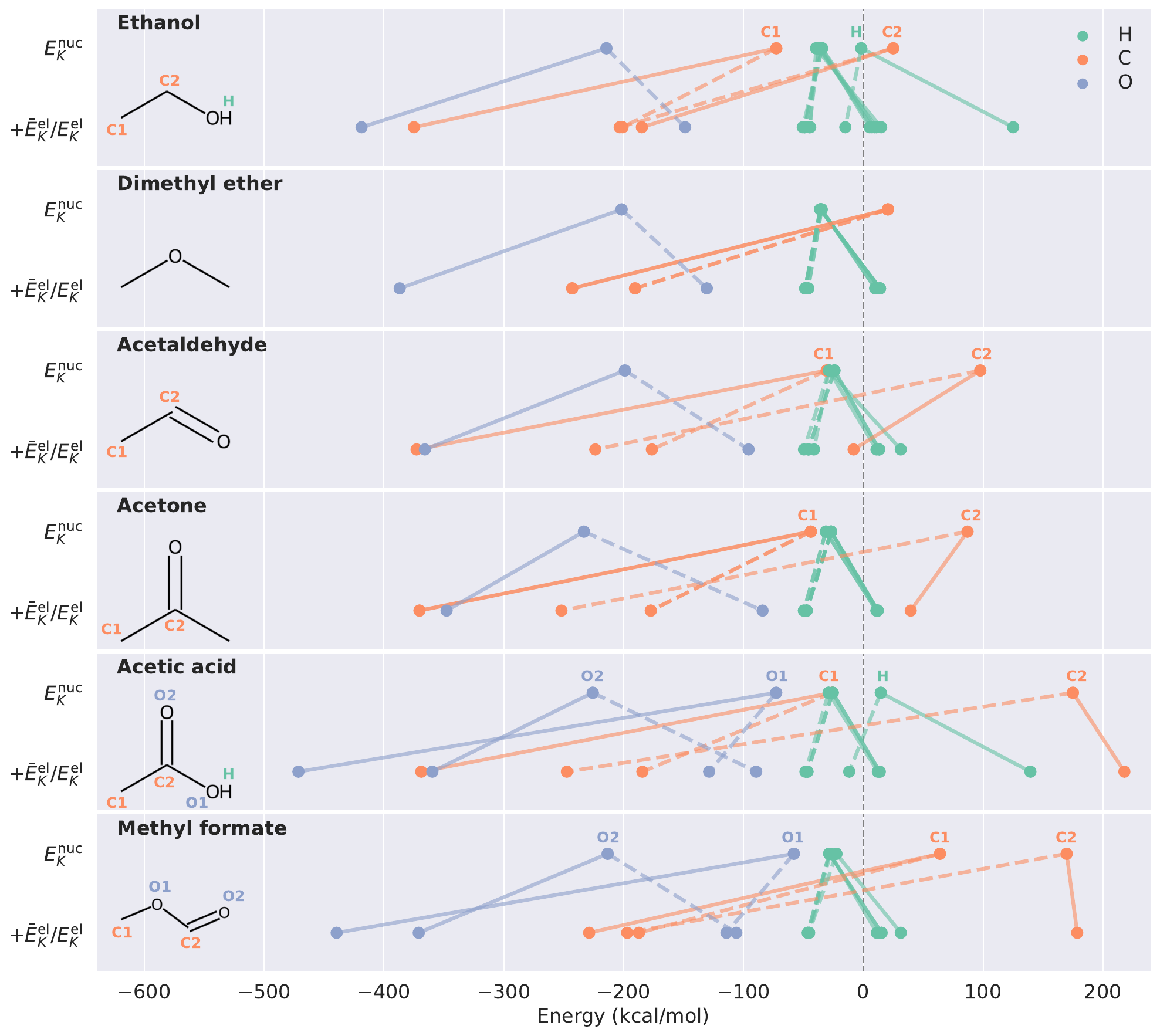}
    \caption{IBO/IAO and EDA results for selected alcohols, ethers, aldehydes, ketones, acetic acids, and esters. For details, please see the caption to Fig. \ref{fig_1}.}
    \label{fig_3}
\end{figure}
\begin{figure}[ht!]
    \centering
    \includegraphics[width=\textwidth]{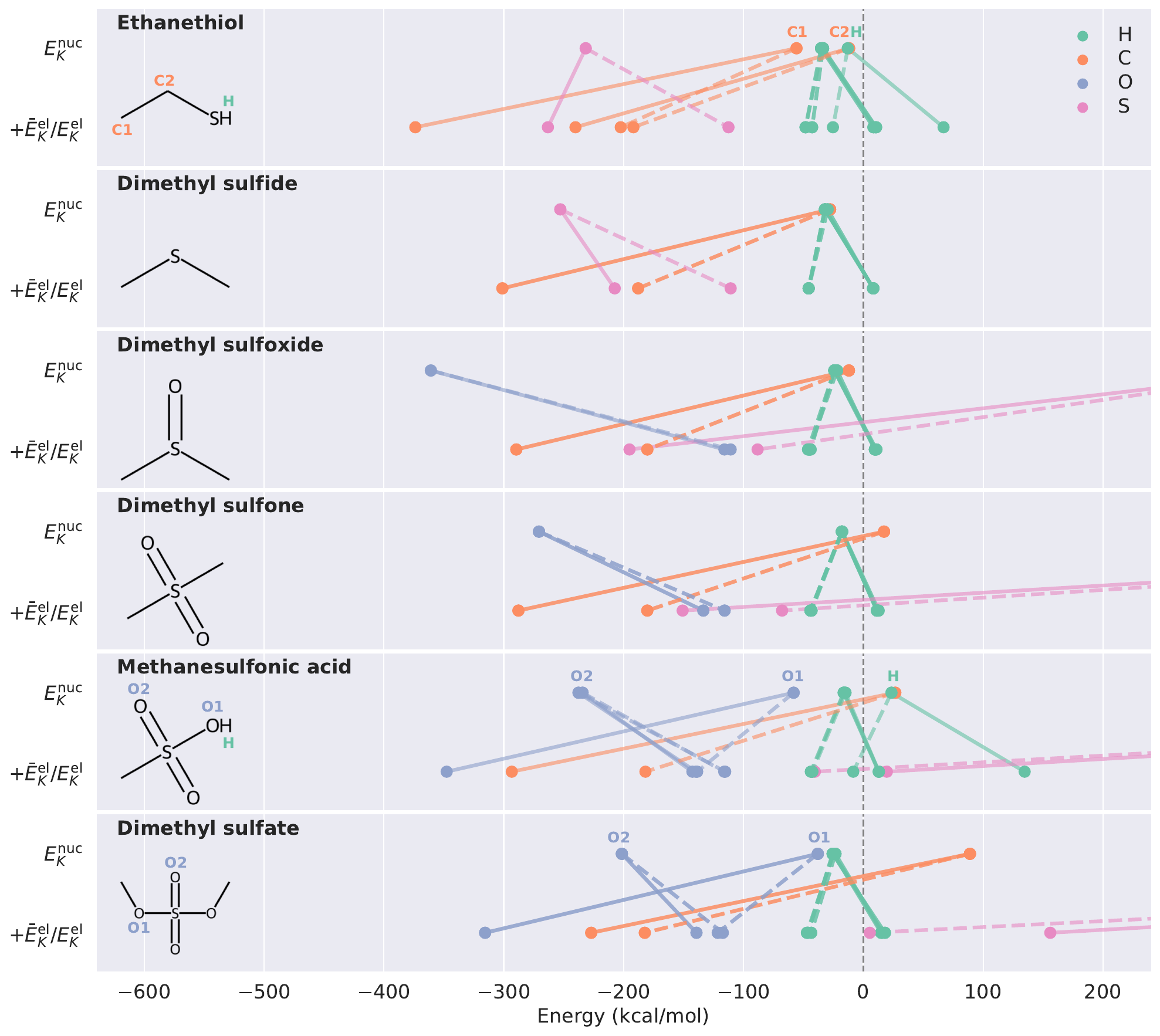}
    \caption{IBO/IAO and EDA results for selected thiols, sulfides, sulfoxides, sulfones, sulfonic acids, and sulfate esters. For details, please see the caption to Fig. \ref{fig_1}.}
    \label{fig_4}
\end{figure}

Fig. \ref{fig_1} presents IBO/IAO decompositions based on Eqs. S1 and S2 and EDA decompositions based on Eqs. S1 and S4 for a selection of six small aliphatic compounds~\bibnote{All results in Figs. \ref{fig_1} through \ref{fig_4} have been computed for geometries derived from corresponding SMILES identifiers, subject only to a simple optimization at the GFN2-xTB level of theory~\cite{grimme_gfn2_xtb_jctc_2019}.}. Focusing initially on the simplest of alkanes in panels 1 through 3, striking contrasts in the description of the local electronic structures around the carbon atoms are observed between the two schemes; the AO-based EDA scheme predicts all constituent carbons to be stabilized in any of the molecules by an equal amount over the free atoms, whilst the MO-based scheme amplifies the stabilization of primary carbons over secondary (propane), tertiary (isobutane), and quaternary (neopentane) carbons, in that same order. As a consequence, only the scheme that builds upon the locality of an MO basis allows for these different carbons to be distinguished from one another in the overall energy spectrum. On account of the dissimilar carbon energies, the contributions associated with the hydrogens also differ between the two schemes. Whereas all hydrogens are stabilized in the EDA scheme, that is, predicted to have a lower energy associated with them in a molecular setting, the MO-based scheme using IBOs and IAOs instead predicts a net, albeit minor destabilization in this case.\\

For the alkenes in panels 4 (propylene) and 5 (isobutylene), the sp$^2$-hybridized carbons are less stabilized than their sp$^3$-hybridized counterparts in the MO-based scheme, again with less stabilization upon an increase in branching. In the AO-based EDA scheme, on the other hand, hardly any distinction is predicted between these different carbons. Finally, for the alkyne in panel 6, both schemes yield notably different contributions. In the scheme based on IBOs/IAOs, the terminal sp-hybridized carbon (C3) is predicted to have a comparable contribution to that of the other member of the triple bond (C2), while this is not the case in the EDA scheme, which nonetheless predicts all hydrogen atoms in the molecule to be nearly identical. This is unlike the MO-based scheme, which instead makes a clear distinction between the hydrogens at either end of the linear hydrocarbon system. The hydrogen (H2) bonded to the sp-hybridized C3 atom is observed to be more destabilized in the molecule than that bonded to the sp$^3$-hybridized C1, corresponding well with the fact that the former type of protons are significantly more acidic than the latter, i.e., $\text{p}K_{\text{a}}\text{(H2)} \ll \text{p}K_{\text{a}}\text{(H1)}$.\\

Turning next to the nitrogen-containing compounds in Fig. \ref{fig_2}, differences between contributions associated with the nitrogens of the three amines in panels 1 through 3 are again more pronounced in the MO-based scheme than in the EDA alternative. As for the alkanes in Fig. \ref{fig_1}, the stabilization predicted by the IBO/IAO scheme diminshes some upon an increase in the number of carbons adjacent to the central nitrogen. For the carbon atoms, the contributions associated with unique classes of these are relatively insensitive with respect to the branching in both schemes, as is evident, e.g., by comparing the contributions from the terminal carbons in ethylamine, the two imines, and the nitrile, all of which have comparable contributions with the corresponding aliphatic carbons in Fig. \ref{fig_1}. However, the MO-based predictions of carbons directly bonded to a nitrogen are observed to differ from the typical values in Fig. \ref{fig_1} on account of the nitrogen neighbour, while this is not the case in the EDA scheme, yielding similar predictions of approx. $-200$ kcal/mol. Likewise, more pronounced distinctions between contributions associated with the hydrogens of the amino/imino groups and those of the methyl group are observed in the MO-based scheme (with more destabilization of the former), arguably due to the dissimilar electronegativities of C and N and the relative differences in local electronic structure around the hydrogens rendered by this.\\

For the imines and nitrile in panels 4 through 6 of Fig. \ref{fig_2}, as touched upon above, the primary carbon contributions of the IBO/IAO scheme are on par with those in Fig. \ref{fig_1} and thus testament to the locality of the electronic structures and their respective energy contributions yielded by this model. On the other hand, the AO-based EDA scheme predicts larger deviations for the terminal carbons and generally stabilizes the carbons bonded to the nitrogen more, a somewhat counter-intuitive observation considering the slightly more unequal sharing of electron density in the $\pi$-bonds between C and N than in the $\sigma$-bonds between the carbons. For acetonitrile in panel 6, the EDA scheme even predicts a destabilization of the nitrogen despite a partial IAO-based Mulliken charge of $q({\text{N}}) = -0.24$ (for reference, $q({\text{C}}2) = +0.14$). This polarization is arguably better reflected in the MO-based contributions where the nitrogen of the cyano group is much more stabilized than the carbon across the triple bond, which, in turn, is much less stabilized than any other carbon of Figs. \ref{fig_1} and \ref{fig_2}.\\

Fig. \ref{fig_3} next presents corresponding results for a number of typical oxygen atoms either singly and doubly bonded to a carbon. For ethanol in the first panel, comparable profiles are observed for our IBO/IAO scheme with the primary amine in Fig. \ref{fig_2}, although the (de)stabilization associated with the hydrogen and oxygen of the alcohol group are exaggerated some with respect to the hydrogens and nitrogen in ethylamine. In the EDA scheme, however, the respective profiles differ in that the electronic response brings about more stabilization of the nitrogen in the amine but less for the oxygen in the alcohol. For the ether in panel 2, a more distinct difference is observed between the stabilization of the oxygen and any of the carbons in the MO-based scheme, but not so in that based on AOs. In both schemes, the stabilization of the oxygen atom is reduced in moving from the alcohol to the ether, a trend which continues upon $\pi$-bonding in the aldehyde, ketone, and carboxlic acid of panels 3 through 5 of Fig. \ref{fig_3}. Unlike in the IBO/IAO scheme, where the carbons of the carbonyl groups become increasingly more destabilized when traversing down these three panels, the EDA scheme predicts these atoms to be increasingly stabilized and even more so than those of the primary methyl groups in these three molecules, again at odds with chemical intuition but also the corresponding IAO-based partial charges; for the carbons denoted as C1 and C2 in panels 3--5 of Fig. \ref{fig_3}, relative differences of $-0.71$, $-0.83$, and $-0.97$, respectively, are predicted in favour of the partially negative terminal carbon (C1).\\

The bottom two panels of Fig. \ref{fig_3} show results for two examples of a carbon (C2) bonded to two different oxygen atoms in two distinct functional groups. In moving from the carboxylic acid to the ester, our MO-based scheme predicts a significant decrease in the stabilization of the terminal methyl carbon (C1, itself bonded to an oxygen in the ester), while the contribution associated with this atom coincides with that of the carbonyl carbon in the EDA scheme, leaving these two atoms pratically indistinguishable from one another in the spectrum of atomization energies. The same holds largely true for the two oxygens in acetic acid and methyl formate in the EDA scheme, whereas the contributions from these are clearly separated in the MO-based scheme. The stabilization of the alcohol oxygen in acetic acid is even predicted to exceed that of the corresponding oxygen in ethanol, true also for the corresponding destabilizations of the hydrogens in these groups (and the lower $\text{p}K_{\text{a}}$ in acetic acid).\\

Finally, as a counterpart to Figs. \ref{fig_1} through \ref{fig_3}, Fig. \ref{fig_4} presents IBO/IAO and EDA results compiled for a selection of prototypical examples in which bonds exist from both carbons and oxygens to central sulfur atoms in a wide range of oxidation states. For both the thiol and sulfide in the first two panels of Fig. \ref{fig_4}, it is obvious how the response of the third-row sulfur atoms to molecular embedding differs from that observed for the nitrogen and oxygen atoms in Figs. \ref{fig_2} and \ref{fig_3}, respectively, while the predictions for the carbon atoms in these molecules align well with all previously considered systems. For instance, the carbon contributions in ethanethiol and dimethyl sulfide coincide almost perfectly with those in ethylamine and dimethylamine of Fig. \ref{fig_2}, true for both the AO- and the MO-based scheme.\\

For those oxygen atoms in the bottom four panels of Fig. \ref{fig_4} that are depicted as doubly bonded, however, our IBO/IAO results are seen to differ from those of oxygens bonded to carbon atoms in Fig. \ref{fig_3}, while this is not true for the results derived from the AO-based EDA scheme. Not only are these oxygen atoms in the IBO/IAO results predicted to be less stabilized upon adding the electronic contribution in Eq. S2---unlike the shift towards more stabilization observed for both the $\pi$-bonded nitrogen and oxygen atoms in Figs. \ref{fig_2} and \ref{fig_3}---their profiles are now also seen to coincide in-between the two schemes. The IBO/IAO profiles of the remaining two oxygens of the sulfonic acid and sulfate ester in panels 5 and 6 of Fig. \ref{fig_4}, respectively, resemble the related profiles in Fig. \ref{fig_3}, but again shifted toward less relative stabilization. Adding to that the fundamentally different profiles (and overall contributions) of the sulfur atoms in question in these molecules, e.g., in comparison with the results in panels 1 and 2, this would appear to indicate that the fundamental character of the involved bonds, at the very least those spanned between sulfurs and oxygens, are peculiarly different from those suspended between the second-row atoms studied in Figs. \ref{fig_1} through \ref{fig_3}. From the canonical Lewis-like depictions of the molecules in Fig. \ref{fig_4}, the central sulfurs also appear to play host to more than eight electrons in their valence spaces, in apparent contradiction with the octet rule~\cite{kutzelnigg_chem_bond_acie_1984}. However, as has been studied by others, compounds like sulfoxides, sulfones, sulfonic acids, and sulfate esters likely observe a dative-type of hypervalency, characterized by highly polarized $\sigma$- rather than $\pi$-type bonds between a central sulfur and its peripheral oxygens~\cite{harshman_miliordos_hypervalency_jce_2020}. To quantify the differences in the bonds studied here between oxygen and either of carbon and sulfur in Figs. \ref{fig_3} and \ref{fig_4}, Table S1 of the SI presents IAO-based Mulliken populations of the involved IBOs (cf. also the similar analysis of sulfur trioxide in Ref. \citenum{knizia_iao_ibo_jctc_2013}). These results further attest to the fact that the oxygens of Fig. \ref{fig_4}, as they are associated with altogether incompatible local electronic structures, need be fundamentally distinguished from those of Fig. \ref{fig_3}. This uniqueness holds true in the results of the MO-based scheme, but not in those based on AOs, as previously alluded to in our recent work on decompositions across small model compound spaces in organic chemistry~\cite{eriksen_qm7_ml_jctc_2023}.\\

Having discussed how AO- and MO-based partitioning schemes yield somewhat contrasting fingerprints of local electronic structures within typical functional groups of organic molecules, we next turn our attention to the different manners in which these schemes decompose changes to atomic properties during the simulation of chemical reactions, e.g., by tracking these along a well-defined reaction coordinate~\cite{morokuma_komornicki_sn2_jcp_1977}. As an initial application, we study the simplest of S$_\mathrm{N}$2 substitution reactions where a halide (nucleophile) replaces a halogen of the same kind on a methyl moiety in a concerted attack from the rear side, i.e., $\mathrm{X}^- \cdots\mathrm{CH}_3\mathrm{X}$ (X = F, Cl, and Br). Here, the simultaneous breaking and formation of two identical bonds occur through a negatively charged transition state (TS) complex in which an aliphatic carbon center is converted from sp$^3$ to approximate sp$^2$ hybridization in a planar, pentacoordinate configuration~\cite{hughes_ingold_sn2_jcs_1937,olmstead_brauman_sn2_jacs_1977,xie_hase_sn2_science_2016}. In accordance with Refs. \citenum{jorgensen_sn2_jacs_1984} and \citenum{jorgensen_sn2_jacs_1985}, the reaction coordinate was therefore defined as $R = R_{\text{C-X(out)}} - R_{\text{C-X(in)}}$ and investigated through a relaxed scan of the distance between C and X(out), starting from the TS. As is well known, the stability of leaving groups in simple gas-phase reactions like these depend primarily on their ability to exist as free anions (i.e., that they are weak bases), but also the strength of the bonds they break to the central carbon atom. This generally makes halides excellent leaving groups, with the notable exception of F owing to how strong bonds between C and F are over those between C and either Cl or Br (with a decreasing trend down the halogens). As fluorine is also exceptionally electronegative, the polarity of the C--F and C--Cl/Br bonds will also differ.\\

\begin{figure}[ht!]
    \centering
    \includegraphics[width=\textwidth]{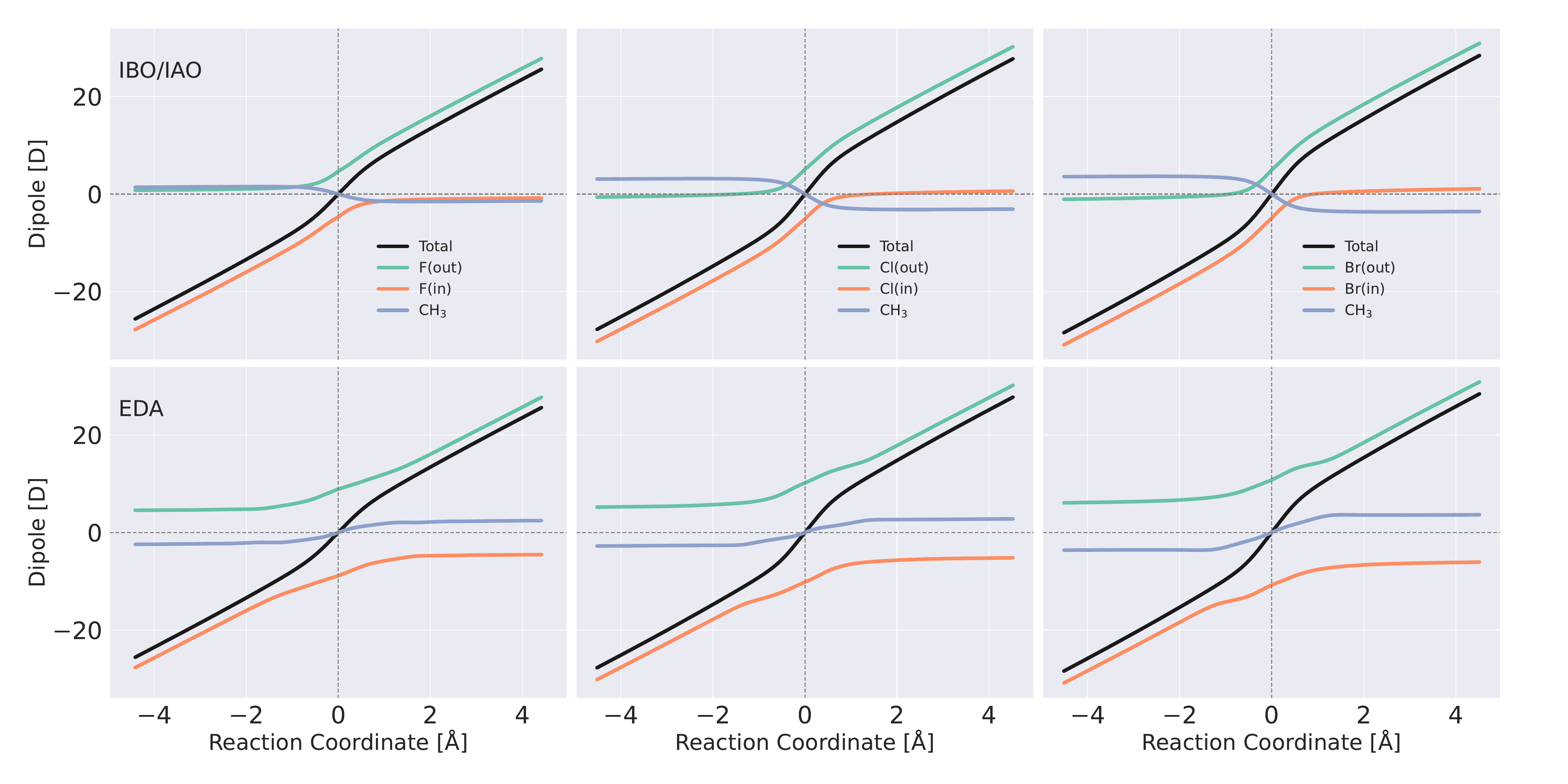}
    \caption{IBO/IAO and EDA atomic dipole moments along each of the studied S$_\mathrm{N}$2 reactions. Results are presented for the non-vanishing dipole component along the $z$-axis only (aligned from positive to negative partial charges), and the gauge origin is chosen to coincide with the position of the central carbon atom at all involved geometries.}
    \label{fig_5}
\end{figure}
We start by comparing the three S$_\mathrm{N}$2 reactions in question by monitoring the response of the molecular dipole moment, as partitioned among the individual atoms, throughout the substitutions (cf. Eqs. S5--S7 of the SI). Due to the total net charge of $-1$, simulations of dipole moments along reactions such as the present will depend on our choice of gauge origin. To ease interpretation, we will set this to coincide with the position of the central carbon atom at all geometries along the reaction coordinate. Using the chemical convention of aligning dipoles from positive to negative partial charges, Fig. \ref{fig_5} compares IBO/IAO and EDA results for the three reactions involving F, Cl, and Br. On the basis of the total dipole moments, all three reactions are practically indistinguishable. Upon decomposing the dipoles, both schemes again yield comparable reaction profiles, albeit with the notable exception of the response of the component associated with the central methyl moiety (i.e., the sum of the vectors assigned to the hydrogens that flip about the carbon center via the planar TS).\\

\begin{figure}[ht!]
    \centering
    \includegraphics[width=\textwidth]{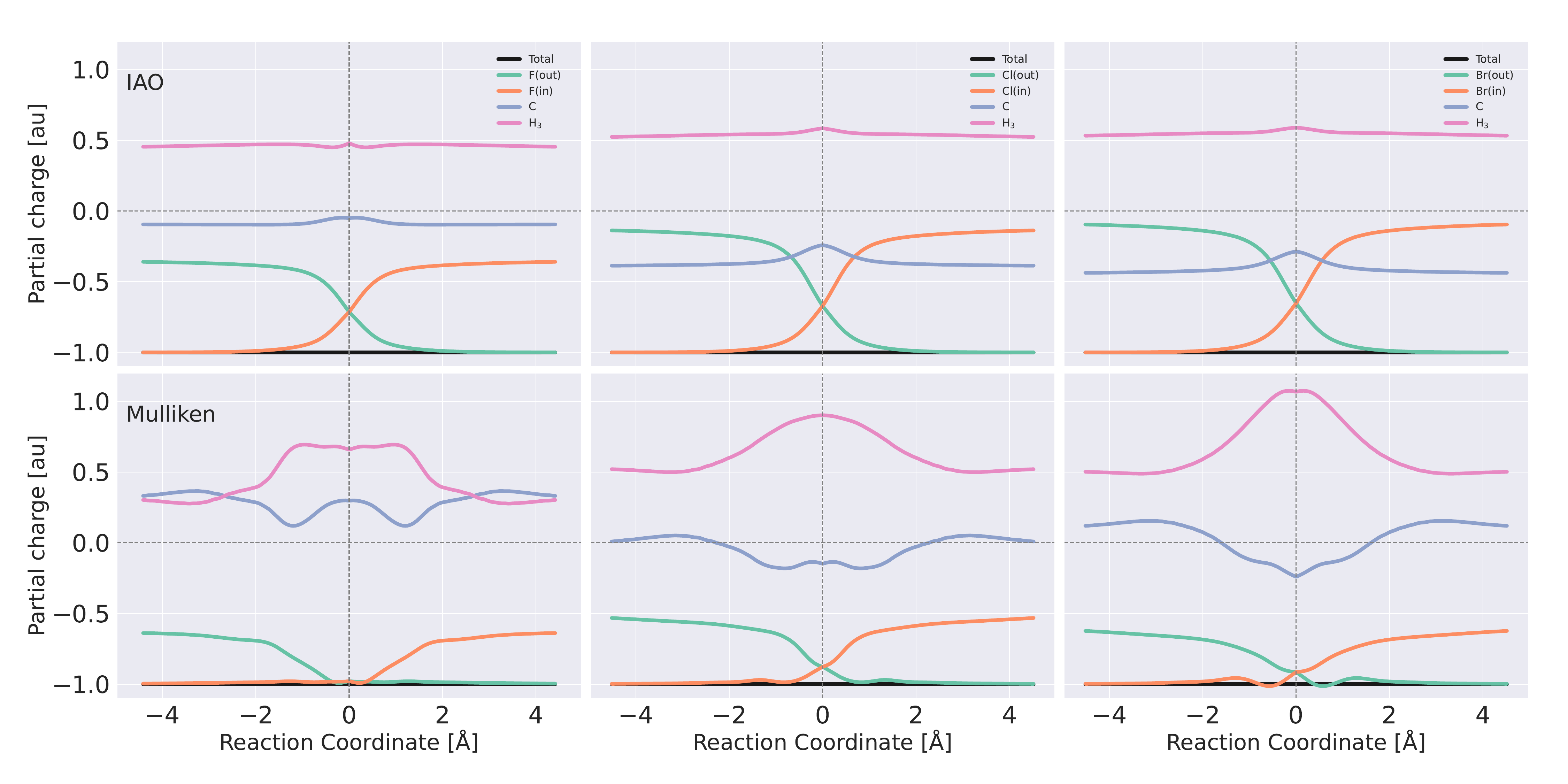}
    \caption{IAO-based and standard Mulliken partial charges along the studied S$_\mathrm{N}$2 reactions.}
    \label{fig_6}
\end{figure}

Whereas the EDA scheme predicts significant contributions associated with each of the bonded halogens, with the accumulated contributions from the hydrogen atoms pointing in the opposite direction, the MO-based IBO/IAO scheme instead paints an altogether different picture with the direction of the latter vector interchanged. To complement the results in Fig. \ref{fig_5}, Fig. \ref{fig_6} compares how the charge distribution changes in each of the three reactions. In contrast to the MO-based IBO/IAO partitioning, the EDA scheme does not explicitly depend on partial populations, but only on the localization of the AO basis. That being said, standard Mulliken charges are derived in the full, computational AO basis, on par with all EDA properties, while IBO/IAO properties are instead computed using a minimal IAO basis, i.e., the exact same intermediate basis used to derive the charges in the upper panel of Fig. \ref{fig_6}.\\

Upon inspection of Fig. \ref{fig_6}, both sets of charges are seen to predict a net transfer of charge to accompany the substitutions. Based on standard Mulliken populations, between $0.3e^{-}$ and $0.4e^{-}$ gets transferred irrespective of the involved halogen, while the IAO populations amplify these charge transfers, predicting $0.6e^{-}$ in the case of flourine and almost a full unit of charge in the substitutions involving chlorine and bromine. As such, all C--X bonds are predicted as equally polarized in the standard Mulliken frame, while the C--F bond clearly differs from those between C and Cl/Br on the basis of the IAO charges. This implies that the level of polarization is thus predicted to be much greater in the full basis than in the IAO counterpart, which can, in turn, be related to the evolution of the atomic dipole moments in Fig. \ref{fig_5} throughout the reactions. In the EDA scheme, the incoming halogen retains most of its excess charge upon formation of its bond to the central carbon, forcing the dipole moment associated with the hydrogens to point in the opposite direction. The central methyl unit is further predicted to be significantly perturbed upon the simultaneous bond formation and cleavage, with the hydrogens taking on more positive partial charge at the expense of the carbon atom. In the IBO/IAO scheme, on the other hand, the balance between partial positive and negative charges in CH$_3$F makes the dipole moments inherent to the molecule much smaller, while the more negative carbon atoms (and comparable hydrogen atoms) render the contributions associated with the methyl moiety relatively larger in the case of CH$_3$Cl and CH$_3$Br. Overall, the perturbations to the methyl moieties in each the reactions are minor and practically negligible in comparison with those predicted by the EDA scheme.\\

\begin{figure}[ht!]
    \centering
    \includegraphics[width=\textwidth]{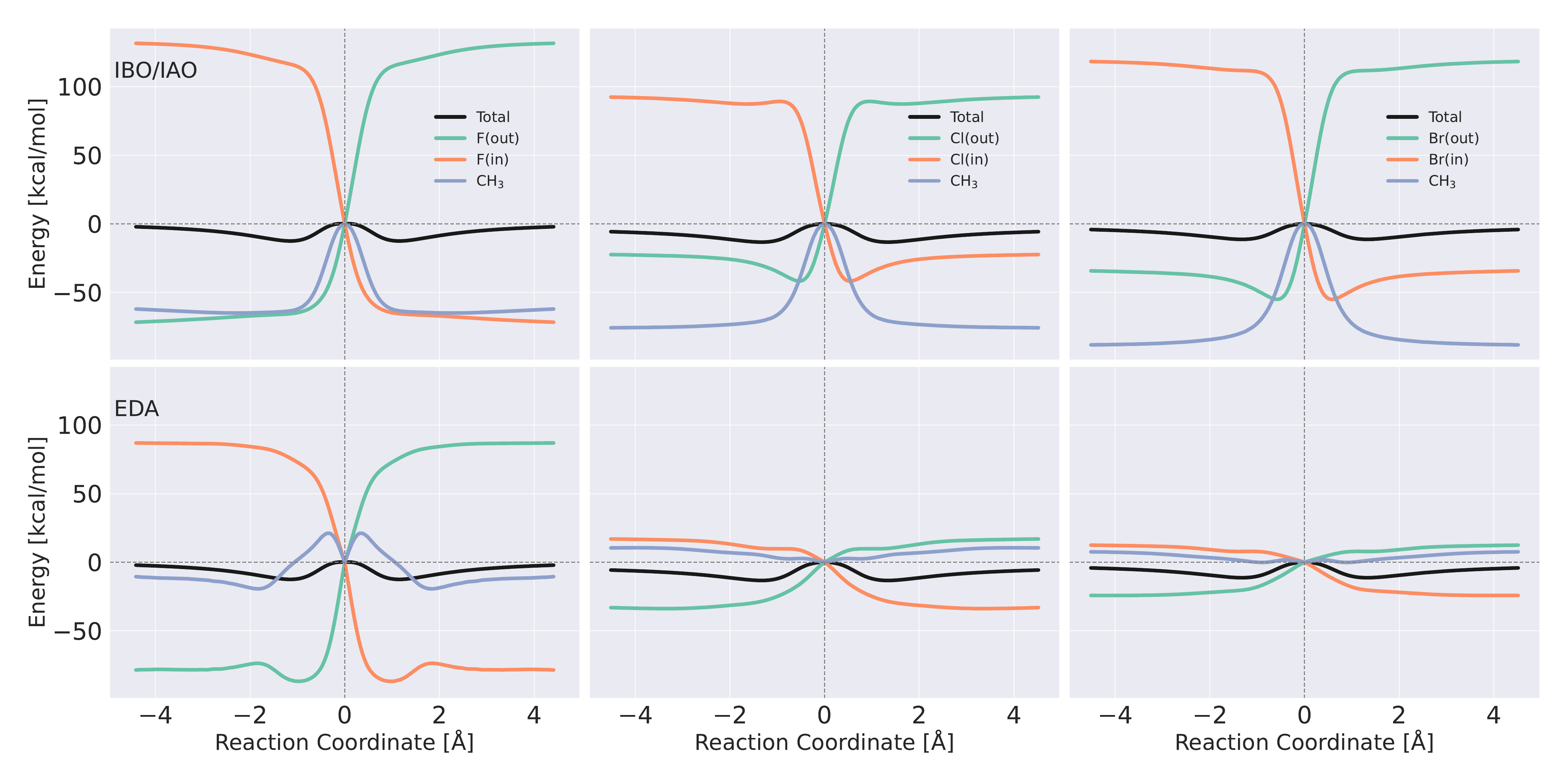}
    \caption{IBO/IAO and EDA atomic energies along each of the studied S$_\mathrm{N}$2 reactions. The energy scale is set such that the atomic energy of each atom is zero at the transition state.}
    \label{fig_7}
\end{figure}
Next, in Fig. \ref{fig_7}, the corresponding changes to individual atomic energies are reported, all relative to the energies at the TS. Again, as in Fig. \ref{fig_5}, the three total reaction profiles depict no real differences between the individual substitutions. The comparable differences in total energies between the separated anions and halomethanes, the ion-dipole stabilized reaction complexes, and the transition states indicate that the mechanisms and driving forces are all but the same in the three reactions. On the other hand, the MO- (IBO/IAO) and AO-based (EDA) partitionings are observed to vary substantially, hinting at intricate, yet fundamental differences in how the substitution reactions proceed at a decomposed, local level.\\

Looking first at the IBO/IAO data, the atomic energies of the central methyl moieties are observed to increase steadily as the reaction progresses, peaking at the TS, before decreasing back down to their initial values, completing the overall Gaussian-like profiles. On the contrary, the atomic energies associated with the incoming halides are predicted to be lowered as the distance to the carbon atom shortens, while the outgoing halogens increase in energy by an equal amount, observing a similar sigmoidal profile. All of these observations are arguably fair in light of what is to be expected, both on the basis of the results on Fig. \ref{fig_1} but also on empirical grounds, in that one bond forms and another simultaneously breaks between two atoms with a substantial difference in electronegativity. However, in the case of fluorine, the lowering in total energy associated with the formation of the reaction complex is seemingly driven by a slight stabilization of the methyl moiety in combination with a continuous stabilization of the incoming fluoride anion. In the case of chlorine and bromine, on the other hand, the stabilization of the reaction complexes is rather attributed to a lowering of the energy of the bonded halogen as the halide approaches from the rear side, while the local environments around the methyl moieties and halides are left practically unperturbed up until this point on the potential energy surfaces. These peculiarities of reactions involving fluorine have also been touched upon by others, e.g., in referring to a type of charge-shift C--F bonding~\cite{shaik_hiberty_cs_bonding_chem_jour_2005}, but previous analyses have arguably not allowed for differences in the evolution of the atomic energies of fluorine and the other halogens to manifest as distinctively along the symmetric substitutions as in the present IBO/IAO results~\cite{schaefer_sn2_jpca_2001,barone_adamo_sn2_jctc_2006,popelier_sn2_jcc_2018}.\\

Turning next to the EDA data in Fig. \ref{fig_7} for comparison, the atomic energies of the central methyl moieties are now predicted to be lowered at the TS. In the case of fluorine, the energy profile for the CH$_3$ unit first passes through a small maximum in-between the reaction complex and the TS, while for chlorine and bromine its stabilization is greatest at the latter point (where the molecular energy is at its maximum), an observation which appears somewhat difficult to rationalize on the basis of chemical intuition alone. Likewise, both the magnitude of the perturbations to the individual atoms in the three reactions, but also the overall profiles are seen to differ significantly when derived from the AO-based scheme, despite similarities in how the total energies change. Combined with the fact that corresponding calculations of atomic dipole moments, partial charges, and energies in a basis set lacking diffuse functions yield fundamentally different results (cf. Figs. S1 through S3 of the SI), the usefulness of native AO-based schemes for interpretative purposes is once more brought into question~\cite{eriksen_qm7_ml_jctc_2023}. For partitioning schemes based on spatially localized MOs, on the other hand, results in both of the (aug-)pc-2 basis sets are indistinguishable, allowing for unbiased interpretations of local electronic structures. Our initial results hence motivate the application of these schemes to even more complex chemistry, whereby finer mechanistic details may be exposed through changes to local rather than total molecular properties.

\section{Summary and Conclusion}

We have investigated the extent to which schemes that partition total KS-DFT properties into a set of intrinsic contributions can yield effective fingerprints indicative of specific types of C, N, O, and S atoms embedded within organic molecules. In particular, we have focused on how atomic contributions to molecular properties from schemes based on the spatial locality of either AOs and MOs align themselves with established chemical intuition, e.g., through the concepts of oxidation states, electronegativity, and charge polarization. In general, partitioning schemes that infer atomic properties solely on the basis of how the underlying AO basis is distributed within a molecular system fail to yield contributions that can distinguish between unique atomic environments. Schemes that instead leverage the locality of suitable sets of MOs are found to successfully express distinct functional groups in physical, organic chemistry, even proving sound and consistent enough to allow for intricate differences to be probed between chemical reactions that are otherwise seemingly uniform in nature.\\

One of the main applications of decomposition schemes like those studied herein is in the mechanistic interpretation of chemical transformations and reactions, not only for elucidating known chemistry, but also for developing guiding principles on how best to resolve electronic structures at an educated theoretical level. In applications to the simplest of S$_\mathrm{N}$2 substitution reactions involving halogens, only schemes formulated in a basis of spatially localized MOs are consistently found to reinforce established chemical intuition and successfully detail the responses of individual atoms along an intrinsic reaction coordinate. In contrast, AO-based partitioning schemes generally tend to predict perturbations to local electronic structures that lend themselves very difficult to interpretation, while further suffering from a strong sensitivity to basis set composition. If one is to relate such derived fingerprints to reactivity proxies in organic chemistry, the heterogeneity of local electronic structures within molecules must be reliably accounted for~\cite{houk_list_aldol_react_jacs_2003,houk_cyclo_add_jacs_2008,coley_qc_aug_nn_jcp_2022}. Computational partitioning schemes formulated in terms of MOs rather than AOs appear the arguably most promising candidates for this purpose.

\section*{Acknowledgments}

This work was supported by two research grants awarded to JJE, no. 37411 from VILLUM FONDEN (a part of THE VELUX FOUNDATIONS) and no. 10.46540/2064-00007B from the Independent Research Fund Denmark.

\section*{Supporting Information}

The supporting information (SI) provides further details on the AO-and MO-based schemes discussed in Sect. \ref{methods_sect}. IAO-based populations of the IBOs in Figs. \ref{fig_3} and \ref{fig_4} are presented in Table S1, and versions of Figs. \ref{fig_5}--\ref{fig_7} computed in the pc-2 basis set are given in Figs. S1--S3.

\section*{Data Availability}

Data in support of the findings of this study are available within the article and its SI.


\providecommand{\latin}[1]{#1}
\makeatletter
\providecommand{\doi}
  {\begingroup\let\do\@makeother\dospecials
  \catcode`\{=1 \catcode`\}=2 \doi@aux}
\providecommand{\doi@aux}[1]{\endgroup\texttt{#1}}
\makeatother
\providecommand*\mcitethebibliography{\thebibliography}
\csname @ifundefined\endcsname{endmcitethebibliography}
  {\let\endmcitethebibliography\endthebibliography}{}

\end{document}